\documentclass[aps,pre,twocolumn,showpacs,preprintnumbers,amsmath,amssymb]{revtex4}

\usepackage{mathrsfs}
\usepackage[dvipdfmx]{graphicx}
\usepackage{dcolumn}
\usepackage{bm}

\newcommand{\mr}{\mathrm}
\newcommand{\mc}{\mathcal}
\newcommand{\tl}{\tilde}
\newcommand{\ra}{\rangle}
\newcommand{\la}{\langle}
\newcommand{\tb}{\textbf}

\begin{document}

\title{Energy pumping in electrical circuits under avalanche noise}

\author{Kiyoshi Kanazawa$^1$, Takahiro Sagawa$^2$, and Hisao Hayakawa$^1$}

\affiliation{
			$^1$Yukawa Institute for Theoretical Physics, Kyoto University, Kitashirakawa-oiwake cho, Sakyo-ku, Kyoto 606-8502, Japan\\
			$^2$Department of Basic Science, The University of Tokyo, Komaba 3-8-1, Meguro-ku, Tokyo 153-8902, Japan
			}
\date{\today}

\begin{abstract}
	We theoretically study energy pumping processes in an electrical circuit with avalanche diodes, where non-Gaussian athermal noise plays a crucial role.
	We show that a positive amount of energy (work) can be extracted by an external manipulation of the circuit in a cyclic way, even when the system is spatially symmetric. 
	We discuss the properties of the energy pumping process for both quasi-static and finite-time cases, and analytically obtain formulas for the amounts of the work and the power. 
	Our results demonstrate the significance of the non-Gaussianity in energetics of electrical circuits. 
\end{abstract}
\pacs{05.70.Ln, 05.10.Gg, 05.40.Fb}

\maketitle
		
\section{Introduction}
	Because of the recent experimental development such as the single molecule manipulation, 
	nonequilibrium statistical mechanics for small systems is a topic of wide interest~{\cite{Bustamante}}. 
	Stochastic thermodynamics~{\cite{SekimotoB,SeifertR,Sekimoto1,Sekimoto2}} in the presence of thermal environment
	has been theoretically studied in terms of nonequilibrium identities~{\cite{Evans,Lebowitz,Crooks,Jarzynski,Kurchan,SeifertIFT,Zon,Noh,Hatano,Harada}}, and 
	is applied to experimental investigations in electrical~{\cite{Ciliberto1,Ciliberto2}} and biological systems~{\cite{Liphardt,Trepagnier,Blickle}}.
	On the other hand, statistical mechanics in the presence of athermal environment has not yet been fully understood, 
	while athermal fluctuation is experimentally known to appear in various systems, such as 
	electrical~{\cite{Gabelli,Zaklikiewicz,Blanter,Onac,Gustavsson}}, biological~{\cite{Brangwynne,Ben-Isaac,Mizuno}}, and granular~{\cite{Talbot,Gnoli}} systems. 
	
	One of the important approaches to athermal statistical mechanics is based on non-Gaussian stochastic models~{\cite{Reimann, Hondou, Luczka, Touchette, Baule, Kanazawa1,Kanazawa2}}, 
	as the crucial property of athermal fluctuation is its non-Gaussianity~{\cite{Gabelli,Brangwynne,Ben-Isaac,Mizuno}}. 
	On the basis of this approach, several interesting phenomena have been reported in athermal systems, 
	which are quite different from thermal ones~{\cite{Kanazawa2,Hondou,Luczka}}. 
	For example, unidirectional transport induced by asymmetric properties of noises or potentials has been discussed with non-Gaussian stochastic models~{\cite{Hondou,Luczka}}. 
	However, there have been so far few studies addressing energy pumping processes of athermal systems. 
	As energy pumping plays crucial roles in thermal physics (i.e., the Carnot cycles~{\cite{Callen,Curzon,Broeck,Schmiedl,Esposito}}), 
	we expect that energy pumping will play important roles in understanding athermal fluctuations. 
	
	In this paper, we study the geometrical pumping~{\cite{Thouless,Berry,Kouwenhoven,Pothier,Brouwer,Breuer,Sinitsyn1,Sinitsyn2,Ohkubo,Ren,Parrondo,SagawaHayakawa,Yuge1,Yuge2}} for athermal systems. 
	When a mesoscopic system is slowly and periodically modulated by several control parameters, there can exist a net average current even without dc bias. 
	This phenomenon is known as the geometrical pumping or the adiabatic pumping, and has been observed in various systems~{\cite{Thouless,Berry,Kouwenhoven,Pothier,Brouwer,Breuer,Sinitsyn1,Sinitsyn2,Ohkubo,Ren,Parrondo,SagawaHayakawa,Yuge1,Yuge2}}.
	The geometrical pumping originates from the effect of the Berry-Sinitsyn-Nemenman phase~{\cite{Berry}}, 
	where a cyclic manipulation in the parameter space induces a nonzero current that is associated with a geometrical quantity on the parameter space. 
	However, all previous studies for open systems address systems connected with thermal or equilibrium reservoirs. 
	Since we encounter athermal systems in various systems, 
	it would be important to study the geometrical pumping coupled with athermal environments. 
	
	Here, we study a realistic geometrical pumping model in an electrical circuit coupled with athermal noise (i.e., avalanche noise). 
	We consider an electrical circuit with a capacitor, resistances, voltages, and avalanche diodes. 
	In the condition with strong reverse voltages, the avalanche diodes produce intermittent fluctuation whose statistics is non-Gaussian~{\cite{Gabelli,Zaklikiewicz}}. 
	We model this system by a non-Gaussian Langevin equation, 
	and find that we can extract a positive amount of work (energy) and power (work per unit time) from the athermal fluctuation as a result of the geometrical effect, while the system is spatially symmetric. 
	We discuss the optimal protocol for the power by using the variational method. 
	Our results show that the athermal fluctuation can be used as an energy source. 
	
	This paper is organized as follows. 
	In Sec. II, we introduce the setup of the electrical circuits with avalanche diodes. 
	In Sec. III, we show the main results of this paper:  the work and power formulas for quasistatic and finite-time processes. 
	In Sec. V, we conclude this paper with some remarks. 
	In Appendix~{\ref{app:manipulation}}, we illustrate an example of the potential manipulation. 
	In Appendix~{\ref{app:derivations}}, we show the detailed derivations of the main results. 
	In Appendix~{\ref{app:arbtrary_potential}}, we generalize our work formula for an arbitrary potential under the condition of a weakly non-Gaussian noise. 
	In Appendix~{\ref{app:integrating_factor}}, we construct a scalar potential for quasistatic work using the method of integrating factors. 
	
\section{System}
	\begin{figure}
		\centering
		\includegraphics[width=75mm]{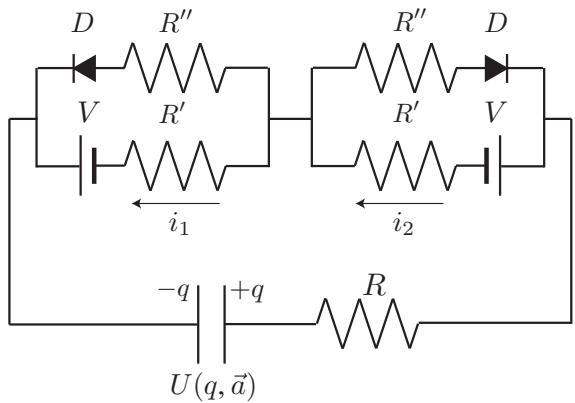}
		\caption{
					Schematic of the electrical circuit with a capacitor with a potential $U(q,\vec a)$, resistances ($R,R',R''$), voltages ($V$), and avalanche diodes ($D$). 
					Because of the reverse bias voltages for the avalanche diodes, the intermittent noise appears and affects the charge in the capacitor. 
				}
		\label{fig:circuit}
	\end{figure}
	We consider an electrical circuit consisting of a capacitor, resistances, avalanche diodes, and external bias voltages (see Fig.~{\ref{fig:circuit}}). 
	Let us denote the charge of the capacitor and time as $q$ and $\bar{t}$, respectively. 
	We note that $\bar{t}$ will be replaced with a scaled time $t$ later. 
	The circuit equation is given by 
	\begin{equation}
		R\frac{dq}{d\bar{t}} + \frac{\partial U(q,\vec a)}{\partial q} - R'i_1 - R'i_2 = 0,
	\end{equation}
	where $R$ and $R'$ are resistances, and $U(q, \vec a)$ is the potential of the capacitor with a set of external parameters $\vec a = (a_1,\dots, a_N)$. 
	It is known that the potential is given by $U(q,d) = \varepsilon_0 A q^2/d$ for a parallel-plate capacitor where $d$, $A$, and $\varepsilon_0$ are, respectively, the width between the plates, the area of the plate, and the vacuum permittivity. 
	Continuous manipulation of the quadratic part of the potential is experimentally realized by changing the width between the plates $d$, 
	where $d$ corresponds to the external parameter as $a_1 = d$ with $N=1$. 
	Nonquadratic potentials can also be realized by inserting a medium with nonlinear permittivity, 
	where we manipulate its nonquadratic part by changing the depth of insertion (see Appendix~{\ref{app:manipulation}} for the details). 
	
	We next discuss the avalanche noise. 
	For sufficiently strong reverse voltages, 
	minority carriers in diodes are accelerated enough to create ionization, producing more carriers which in turn create more ionization.
	Thus, electrical current is multiplied to become an intermittent noise. 
	This noise is known as the avalanche noise, which can be approximated as a white non-Gaussian noise in the case of a high level of avalanche~{\cite{Gabelli, Zaklikiewicz}}. 
	When we decompose $i_n$ into the steady and fluctuating parts as $i_n=\langle i_n\rangle + \Delta i_i$ for $n=1,2$,
	$\Delta i_n$ can be regarded as a white non-Gaussian noise. 
	In the following, $\langle{A}\rangle$ denotes the ensemble average of a stochastic variable $ {A}$, and the Boltzmann constant is taken to be unity. 
	Then, the time evolution of the charge in the capacitor is reduced to the following Langevin equation: 
	\begin{equation}
		\frac{dq}{dt} = -\frac{\partial U(q, \vec a)}{\partial q} + \xi,\label{eq:Langevin_NG}
	\end{equation}
	where $t \equiv \bar{t}/(R+2R')$ is the scaled time, and $\xi\equiv R'(\Delta i_1 + \Delta i_2)$ is the white non-Gaussian noise which describes the avalanche noise. 
	Because of the bilateral symmetry in the circuit, we assume that $\xi$ is symmetric for the charge reversal. 
	We stress that similar Langevin equations to Eq.~{(\ref{eq:Langevin_NG})} appear in many mesoscopic systems, such as electrical circuits with shot noises~{\cite{Blanter,Gardiner}} and ATP-driven active matters~{\cite{Brangwynne,Ben-Isaac}}. 
	Therefore, it is straightforward to apply our formulation to a wide class of mesoscopic systems beyond the electrical circuit addressed in this paper. 
	The cumulants of the noise are given by 
	\begin{equation}
		\langle \xi(t_1)\dots \xi(t_{n})\rangle_c= 	\begin{cases}
														K_{n}\delta_{n}(t_1,\dots,t_{n}) & {\rm (for\>\> even\>\>}n)\cr
														0 & {\rm (for\>\> odd\>\>}n)
													\end{cases},
	\end{equation}
	where $\langle \xi(t_1)\dots \xi(t_n)\rangle_c$ denotes the $n$th cumulant,
	and $\delta_n(t_1,\dots,t_n)$ is an $n$-point $\delta$ function~\cite{Kleinert,Kanazawa2} with a positive integer $n$.
	We note that the $n$-point $\delta$ function satisfies the following relations as 
	\begin{align}
		\delta_n (t_1,\dots,t_n) &= \begin{cases}
										\infty & (t_1=\dots=t_n)\\
										0 & (\mathrm{otherwise})
									\end{cases},\\
		\int_{-\infty}^{\infty}dt_2\dots dt_n &\delta_n(t,t_2,\dots t_n) = 1,
	\end{align}
	where we introduce $T\equiv K_2/2$ for later convenience. 
	To extract work, we externally manipulate this system through a cyclic operational protocol $C\equiv \{\vec a(t)\}_{0\leq t\leq \tau}$, 
	where $\tau$ is the period of the manipulation, and the cyclic protocol satisfies the relation as $\vec a(0)=\vec a(\tau)$. 
	On the basis of stochastic energetics~{\cite{SekimotoB,Sekimoto1,Sekimoto2}}, we define the extracted work $W$ as 
	\begin{equation}
		d  W \equiv -\frac{\partial U}{\partial \vec a}\cdot d\vec a = -\sum_{i=1}^N\frac{\partial U}{\partial a_i}da_i. \label{eq:work}
	\end{equation} 
	In the special case of $K_{n}=0$ for $n\geq4$, the Langevin equation~{(\ref{eq:Langevin_NG})} is equivalent to the thermal Gaussian Langevin equation, 
	and we cannot extract positive work from the fluctuation~\cite{SekiSasa,SekimotoB}:  
	\begin{equation}
		\oint_{C} dW_{\mr{qs}}\leq  0,
	\end{equation}
	where the equality holds for the quasistatic processes. 
		
\section{Main results}
	In this section, we discuss the main results of this paper: the formulas for the work and the power of the geometrical pumping from athermal fluctuations. 
	\subsection{Work along quasistatic processes}
\begin{figure}
			\centering
			\includegraphics[width=80mm]{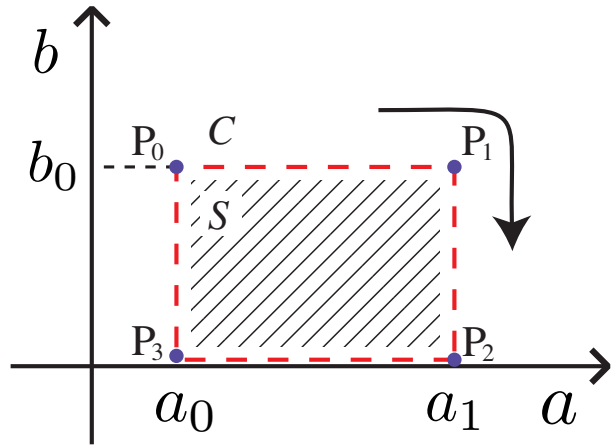}
			\caption{
						(Color online) Schematic of the rectangular protocol. 
						We assume $a_0=O(1)$, $a_1=O(1)$, $a_1-a_0 = O(1)$, and $b_0=O(\epsilon)$.
						We can extract a positive amount of work from the nonequilibrium fluctuation along the clockwise protocol. 
					}
			\label{fig:protocol}
		\end{figure}

		First of all, we consider a weakly quartic potential 
		\begin{equation}
			U(q,\vec a) = \frac{aq^2}{2} + \frac{bq^4}{4},\label{eq:potential}
		\end{equation} 
		where $\vec a=(a,b)$ are two external parameters. 
		We also assume that $b$ is proportional to a small parameter $\epsilon$.  
		We then obtain, for quasistatic processes,
		\begin{equation}
			dW_{\mr{qs}} = -d\left(\frac{T}{2}\log{a} + \frac{3bT^2}{4a^2} + \frac{bK_4}{16a}\right) + \frac{bK_4}{16a^2}da+ O(\epsilon^2), \label{eq:wqp_formula}
		\end{equation}
		which will be proved in Appendix~{\ref{app:derivations}}. 
		Equality~{(\ref{eq:wqp_formula})} implies that there exists a quasi-static cyclic protocol $C_{\mr{qs}}$ along which a positive amount of work can be extracted as 
		\begin{equation}
			W_{\mr{qs}}\equiv \oint_{C_\mr{qs}} dW_{\mr{qs}} = \oint_{C_{\mr{qs}}} \frac{bK_4}{16a^2}da > 0, \label{eq:circ_work}
		\end{equation}
		even though the potential and the noise are spatially symmetric throughout the control protocol. 
		For example, a positive amount of work can be extracted through the clockwise rectangular protocol (Fig.~{\ref{fig:protocol}}) as $W_{\mr{qs}} = (bK_4/16)[1/a_0-1/a_1]$. 
		In Eq.~{(\ref{eq:wqp_formula})}, the fourth-order cumulant appears because the perturbative potential is quartic. 
		If the perturbative potential includes another higher-order polynomial, the corresponding order cumulants appear as correction terms.  
		We note that our result does not contradict the second law of thermodynamics, because the avalanche noise is nonequilibrium fluctuation 
		(i.e., the environment is out of equilibrium). 
		We also note that the work formula~{(\ref{eq:wqp_formula})} for quasistatic processes can be extended for an arbitrary potential 
		for weakly non-Gaussian cases (see Appendix~\ref{app:arbtrary_potential} for detail).

		The pumping effect in Eqs.~{(\ref{eq:wqp_formula})} and {(\ref{eq:circ_work})} can be regarded as the geometrical effects of the Berry-Sinitsyn-Nemenman phase~{\cite{Thouless,Berry,Kouwenhoven,Pothier,Brouwer,Breuer,Sinitsyn1,Sinitsyn2,Ohkubo,Ren,Parrondo,SagawaHayakawa,Yuge1,Yuge2}}. 
		Indeed, by introducing $\chi\equiv -(T/2)\log{a}-3bT^2/4a^2-bK_4/16a$, $\vec{\mathcal{A}} \equiv (bK_4/16a^2,0)$, $\Omega\equiv K_4/16a^2$, and $S_{\mr{qs}}$ (the area surrounded by $C_{\mr{qs}}$),
		we can rewrite Eqs.~{(\ref{eq:wqp_formula})} and {(\ref{eq:circ_work})} as 
		\begin{equation}
			dW_{\mr{qs}} = d\chi + \vec{\mathcal{A}}\cdot d\vec a + O(\epsilon^2),
		\end{equation}
		\begin{equation}
			\oint_{C_\mr{qs}} dW_{\mr{qs}} = \oint_{C_{\mr{qs}}} \vec{\mathcal{A}}\cdot d\vec a = \int_{S_{\mr{qs}}} \Omega dadb. 
		\end{equation}
		This expression implies that $\chi$, $\vec{\mathcal{A}}$, and $\Omega$ respectively correspond to 
		the scalar potential, the vector potential, and  the curvature in the terminology of the Berry phase. 
		We note that the curvature $\Omega$ is nonzero since $dW_{\rm qs}$ is an inexact differential, which creates a nonzero geometrical pumping current for cyclic operations. 
		
		We remark on the relation between thermodynamic scalar potentials and the method of integrating factors. 
		In the presence of thermal environments, the integrated quasistatic work $\Delta F = \int dW_{\mr{qs}}$ is the thermodynamic scalar potential (Helmholtz's free energy). 
		On the other hand, in athermal cases, $\int dW_{\mr{qs}}$ is no longer regarded as a scalar potential because of the presence of the nonzero curvature. 
		Even in such situations, the method of integrating factors is useful to find a scalar potential if it exists, 
		because the integrating factors can make an inexact differential an exact differential. 
		We stress that we find an explicit integrating factor if we focus on the case with the weakly quartic potential as shown in Appendix~\ref{app:integrating_factor}, 
		though there are not necessarily appropriate integrating factors for general athermal cases. 
		
\begin{figure}
			\centering
			\includegraphics[width=80mm]{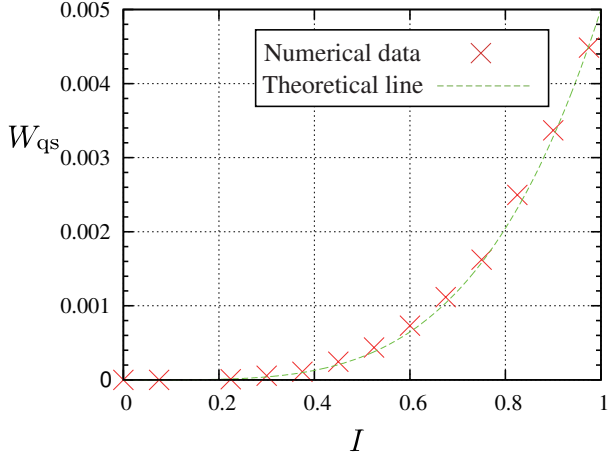}
			\caption{
						(Color online) Numerical validation of the work formula~{(\ref{eq:wqp_formula})} for the quasistatic processes. 
						From the Monte Carlo simulation, we obtain stochastic trajectories and calculate the ensemble average of the extracted work. 
						We calculate the work with the total time of the operation $\tau = 3.0 \times 10^4$
						and take its ensemble average with $6600$ samples. 
						Here we assume the discretized time step is $10^{-2}$. 
						The time scaled protocol for the simulation $(\tl{a}(\tl{s}),\tl{b}(\tl{s}))\equiv (a(\tau\tl{s}),b(\tau\tl{s}))$ is given as follows: 
						$\tl{a}(\tl{s})=a_1\>(0\leq \tl{s}\leq 1/4), 4a_1(1/2-\tl{s}) + 2a_0(\tl{s}-1/4)\>(1/4\leq \tl{s}\leq 1/2),
						a_0\>(1/2\leq \tl{s}\leq 3/4), 4a_1(\tl{s}-3/4) + 4a_0(1-\tl{s})\>(3/4\leq \tl{s}\leq 1)$ and 
						$\tl{b}(\tl{s})=4b_0(1/4-\tl{s})\>(0\leq \tl{s}\leq 1/4), 0\>(1/4\leq \tl{s}\leq 1/2),
						4b_0(\tl{s}-1/2)\>(1/2\leq \tl{s}\leq 3/4), b_0\>(3/4\leq \tl{s}\leq 1)$.
					}
			\label{fig:work_qs}
		\end{figure}

		We numerically check the validity of Eqs.~{(\ref{eq:wqp_formula})} and {(\ref{eq:circ_work})} by the Monte Carlo simulation. 
		For simplicity, we model the avalanche noise as the symmetric Poisson noise defined by 
		\begin{equation}
			  \xi_{S}(t) = \sum_{i=0}^\infty I\delta(t-  t_i) + \sum_{i=0}^\infty (-I)\delta(t-  s_i),\label{eq:symmetric_Poisson}
		\end{equation}
		where $  t_i$ and $  s_i$ are times where the Poisson flights happen with the flight distance $\pm I$ and the transition rate $\lambda/2$. 
		We note that the cumulants are given as $2T=I^2\lambda$ and $K_{2n}=I^{2n}\lambda$ with integer $n\geq 2$.
		We consider a rectangular protocol shown in Fig.~{\ref{fig:protocol}} and set parameters as $a_0=1.0$, $a_1=5.0$, $b_0=0.1$, and $\lambda=1.0$. 
		Changing the flight distance parameter $I$, we numerically obtain the work for the rectangular quasistatic protocol.  
		Figure~{\ref{fig:work_qs}} shows that the numerical results are consistent with the theoretical line obtained in Eq.~{(\ref{eq:wqp_formula})}. 
		This result implies that we can extract more energy from the athermal fluctuation as the non-Gaussian property characterized by the flight distance $I$ increases. 

	\subsection{Power along slow operational processes}
		We next consider the power of the energy pumping for the weakly quartic potential~{(\ref{eq:potential})}. 
		Let $C$ be a cyclic protocol of the operation in the $a$-$b$ space and $\tau$ be the total time of the operation. 
		We introduce time-scaled external parameters $\tl{a}(\tl{s})$, $\tl{b}(\tl{s})$ and a time-scaled protocol $\tl{C}\equiv \{\tl{a}(\tl{s}),\tl{b}(\tl{s})\}_{0\leq \tl{s}\leq 1}$, 
		where $\tl{a}(\tl{s})$ and $\tl{b}(\tl{s})$ are scaled by the total operational time $\tau$ as $\tl{a}(\tl{s}) \equiv a(\tau\tl{s})$ and $\tl{b}(\tl{s})\equiv b(\tau\tl{s})$. 
		Because we are interested in slow but finite-time processes, we assume that $1/\tau$ is the order of $\epsilon$, $d\tl{a}/ds = O(1)$, and $d\tl{b}/ds = O(\epsilon)$. 	
		As will be shown in Appendix~{\ref{app:derivations}} with a similar calculation to that in Ref.~{\cite{SekiSasa}}, the work for slow operational processes is given by 
		\begin{align}
			\int \langle d  W\rangle &= \int dW_{\mr{qs}} - \frac{1}{\tau}S[\tl{C}] + O(\epsilon^2),\label{eq:wqp_formula_fs}\\
			S[\tl{C}] &= \int_0^1 \frac{d\tl{s}T}{4\tl{a}^3}\left[\frac{d\tl{a}}{d\tl{s}}\right]^2.\label{eq:wqp_formula_fs2}
		\end{align}
		From Eq.~{(\ref{eq:wqp_formula_fs})}, we obtain the average power: 
		\begin{equation}
			P\equiv \frac{1}{\tau}\oint_C \langle d  W\rangle 
			= \frac{1}{\tau}\oint_{C_{\mr{qs}}} \frac{bK_4}{16a^2}da
			- \frac{1}{\tau^2}S[\tl{C}] + O(\epsilon^3).\label{eq:wqp_formula_pwr}
		\end{equation}
		The optimal total time that maximizes the power under a fixed time-scaled protocol $\tl{C}$ is derived from the condition
		\begin{equation}
			\frac{dP}{d\tau}\bigg|_{\tau=\tau^*}=-\frac{1}{\tau^2}\oint_{C_{\mr{qs}}} \frac{bK_4}{16a^2}da+ \frac{2}{\tau^3}S[\tl{C}]=0,
		\end{equation}
		which leads to
		\begin{equation}
			\tau^* \equiv \frac{2S[\tl{C}]}{\oint_{C_{\mr{qs}}} (bK_4/16a^2)da}. \label{eq:optimal_time}
		\end{equation}
		We note that Eq.~{(\ref{eq:optimal_time})} is consistent with the assumption $\tau=O(1/\epsilon)$. 
		Thus we obtain the optimal power for the fixed scaled protocol as 
		\begin{equation}
			P^* \equiv \frac{\left[\oint_{C_{\rm qs}}(bK_4/16a^2)da\right]^2}{4S[\tl{C}]} + O(\epsilon^3).
		\end{equation}
		
\begin{figure}
			\centering
			\includegraphics[width=85mm]{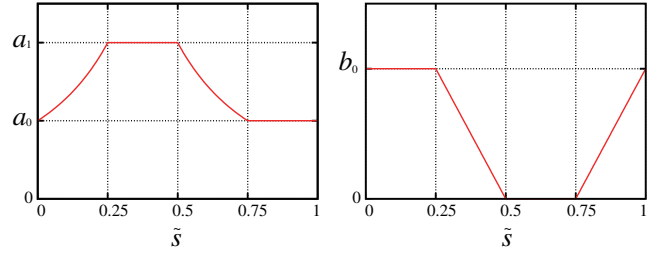}
			\caption{	(Color online) Scaled optimal rectangular protocol~{(\ref{eq:opt_prtcl1})} and (\ref{eq:opt_prtcl2})  
						on the condition of $\tl{a}(0)=\tl{a}(3/4)=\tl{a}(1)=a_0$, $\tl{a}(1/4)=\tl{a}(1/2)=a_1$, $\tl{b}(0)=\tl{b}(1/4)=\tl{b}(1)=b_0$, and $\tl{b}(1/2)=\tl{b}(3/4)=0$.}
			\label{fig:opt_protocol}
		\end{figure}

		As an example, let us consider the rectangular protocol shown in Fig.~{\ref{fig:protocol}}, 
		where the manipulation proceeds as $\rm P_0 \rightarrow P_1 \rightarrow P_2 \rightarrow P_3\rightarrow P_0$. 
		We denote the arrival time for P$_i$ as $t_i$ for $i=1,2,3$, and rescale $t_i$ as $\tl{\tau}_i\equiv t_i/\tau$. 
		We assume that $\tl{\tau}_i=i/4$ for $i=1,2,3$, where $d\tl{a}/ds = O(1)$ and $d\tl{b}/ds = O(\epsilon)$ are satisfied.
		We then consider the optimal protocol for the rectangular protocol.  
		We explicitly obtain
		\begin{equation}\label{eq:ineq_opt}
			S[\tl{C}] \geq 8T\bigg|\frac{1}{\sqrt{a_0}} - \frac{1}{\sqrt{a_1}}\bigg|^2,
		\end{equation}
		which will be proved in Appendix~{\ref{app:derivations}}. 
		Here, the equality holds for the optimal scaled protocol $\tl{C}_{\rm opt}\equiv \{\tl{a}^*(\tl{s}), \tl{b}^*(\tl{s})\}_{0\leq \tl{s} \leq 1}$ given by (see Fig.~{\ref{fig:opt_protocol}})
		\begin{align}
			\tl{a}^*(\tl{s}) &= 
					\begin{cases}
						\big|\frac{4\tl{s}}{\sqrt{a_1}}+\frac{1-4\tl{s}}{\sqrt{a_0}} \big|^{-2} & (0\leq \tl{s} \leq \frac{1}{4})\cr
						a_1 & (\frac{1}{4} \leq \tl{s} \leq \frac{1}{2})\cr
						\big|\frac{3-4\tl{s}}{\sqrt{a_1}}+\frac{4\tl{s}-2}{\sqrt{a_0}} \big|^{-2} & (\frac{1}{2} \leq \tl{s} \leq \frac{3}{4})\cr
						a_0 & (\frac{3}{4} \leq \tl{s} \leq 1)
					\end{cases},\label{eq:opt_prtcl1}\\
			\tl{b}^*(\tl{s}) &= 
					\begin{cases}
						b_0   & (0\leq \tl{s} \leq \frac{1}{4})\cr 
						2b_0(1-2\tl{s})   & (\frac{1}{4} \leq \tl{s} \leq \frac{1}{2})\cr 
						0   & (\frac{1}{2} \leq \tl{s} \leq \frac{3}{4})\cr
						b_0(4\tl{s}-3)   & (\frac{3}{4} \leq \tl{s} \leq 1)
					\end{cases}.\label{eq:opt_prtcl2}
		\end{align}
		We then obtain the maximum power as
		\begin{equation}
			P^* = \frac{1}{2T}\left[\frac{bK_4}{64}\right]^2\left|\frac{1}{\sqrt{a_0}}+\frac{1}{\sqrt{a_1}}\right|^2 + O(\epsilon^3). \label{eq:power}
		\end{equation}
		This result exhibits that a positive amount of power is extracted from the avalanche noise as the non-Gaussianity increases. 
		The optimal total time of the operation is given by 
		\begin{equation}
			\tau^* = \frac{256T}{bK_4}\frac{1/\sqrt{a_0}-1/\sqrt{a_1}}{1/\sqrt{a_0}+1/\sqrt{a_1}}.\label{eq:opttime}
		\end{equation}
		We have some remarks on the validity of Eqs.~{(\ref{eq:opt_prtcl1})}, {(\ref{eq:opt_prtcl2})}, and (\ref{eq:power}).
		According to Eq.~{(\ref{eq:wqp_formula_pwr})}, the processes $\rm P_1\rightarrow P_2$ and $\rm P_3 \rightarrow P_0$ are irrelevant for $S[\tl{C}]$.
		Therefore, the explicit form of Eq.~{(\ref{eq:opt_prtcl2})} is arbitrary for $1/4\leq \tl{s} \leq 1/2$ and $3/4\leq \tl{s} \leq 1$ 
		if the following assumptions are satisfied: $\tl{b}(1/4)=b_0$, $\tl{b}(1/2)=0$, $\tl{b}(3/4)=0$, $\tl{b}(1)=b_0$, and $d\tl{b}/d\tl{s}=O(\epsilon)$. 
		We also note that the formula~{(\ref{eq:power})} is only valid under the assumptions of $a_0=O(1)$, $a_1=O(1)$, and $a_1-a_0=O(1)$, 
		which implies that Eq.~{(\ref{eq:power})} is invalid for some limits such as $a_0-a_1\rightarrow +0$ or $a_1\rightarrow \infty$.
		
\begin{figure}
			\centering
			\includegraphics[width=83mm]{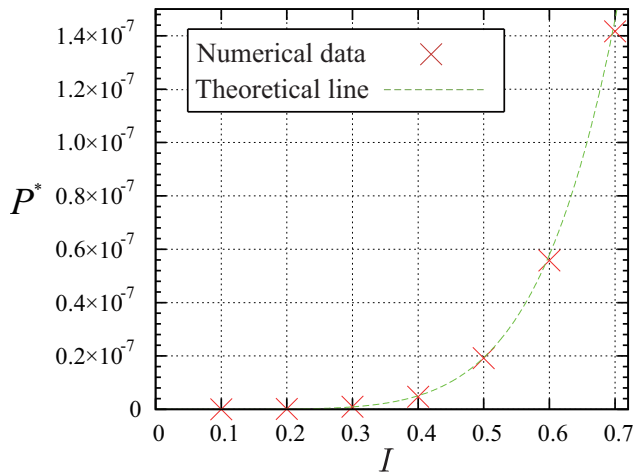}
			\caption{
						(Color online) Numerical demonstration of the validity of the power formula~{(\ref{eq:power})}. 
						On the basis of the method of Monte Carlo, we numerically obtain trajectories 
						with the fourth Runge Kutta method and take the ensemble average of the extracted power with the discretized time step as $\Delta t=0.005$. 
						The ensemble number depends on the parameter $I$. For example, the ensemble number is approximately equal to $1.14 \times 10^7$ for $I=0.7$.
					}
			\label{fig:nmrcl_demo_power}
		\end{figure}

		We numerically verify the validity of the power formula~{(\ref{eq:power})} for the rectangular optimal protocol~{(\ref{eq:opt_prtcl1})}, {(\ref{eq:opt_prtcl2})}, and {(\ref{eq:opttime})}. 
		We consider the symmetric Poisson model~{(\ref{eq:symmetric_Poisson})} on the condition that $a_0=1$, $a_1=5$, $b_0=0.05$, and $\lambda = 1.0$. 
		We control the flight distance $I$, and we plot the average power as a function of $I$ in Fig.~{\ref{fig:nmrcl_demo_power}}. 
		The numerical data in Fig.~{\ref{fig:nmrcl_demo_power}} are consistent with the theoretical line~{(\ref{eq:power})}, 
		which implies that a more positive amount of power is extracted by this engine as the non-Gaussianity increases. 

\section{Concluding remarks}
	We have studied the energy pumping of an electrical circuit consisting of avalanche diodes. 
	Using this circuit, we can extract a positive amount of work from the non-equilibrium fluctuations of the avalanche diodes 
	even though the fluctuation and the potential are spatially symmetric. 
	We derive the work and power formulas~{(\ref{eq:wqp_formula})} and~{(\ref{eq:wqp_formula_pwr})} to discuss quasistatic and finite-time operational processes. 
	We have checked the validity of our formulas through numerical simulations. 
	Our theory can be used to measure high-order cumulants of the avalanche noise. 

	We remark that our formulation would be applicable to other athermal systems, 
	such as granular~{\cite{Talbot,Gnoli}} and biological~{\cite{Mizuno}} systems. 
	For example, if we regard the charge in the capacitor as the angle of the granular motor, 
	the circuit corresponds to the motor driven by the dilute granular gas with the air friction. 
	It is also interesting to generalize our formulation for non-Markovian systems. 
	
\begin{acknowledgements}
	We gratefully acknowledge K. Chida and H. Takayasu for detailed discussion on experimental realization. 
	We also thank T. G. Sano, S. Ito, F. van Wijland, P. Visco, and \'E. Fodor for valuable discussions. 
	A part of the numerical calculations was carried out on SR16000 at YITP in Kyoto University. 
	This work was supported by the JSPS Core-to-Core Program ``Non-equilibrium dynamics of soft matter and information," 
	the Grants-in-Aid for Japan Society for Promotion of Science (JSPS) Fellows (Grant No. 24$\cdot$3751), 
	and JSPS KAKENHI Grants No. 22340114 and No. 25800217. 
\end{acknowledgements}

\appendix
	\section{A possible example of the potential manipulation}\label{app:manipulation}
\begin{figure}
			\includegraphics[width=80mm]{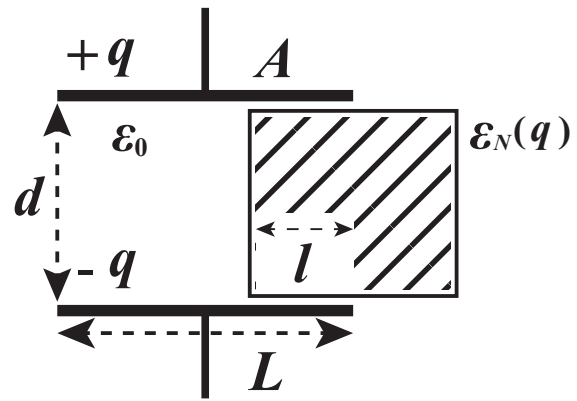}
			\caption{	Schematics of the potential manipulation by inserting a medium with nonlinear permittivity. 
						The medium with nonlinear permittivity $\varepsilon_N(q)$ is inserted as shown in this figure 
						to control the potential of the capacitor. 
					}
			\label{fig:manipulation}
		\end{figure}

		In this appendix, we illustrate a possible example to realize the potential manipulation using medium with nonlinear permittivity. 
		Let us consider a capacitor composed of two parallel plates with their area $A$ and distance $d$ as shown in Fig.~{\ref{fig:manipulation}}. 
		We externally insert a medium with the nonlinear permittivity $\varepsilon_N(q)$ into the space between the plates. 
		Let us denote the insertion depth of the medium by $l$. 
		Then, the potential of the capacitor can be written as 
		\begin{equation}
			U(q,d,r) = \frac{Aq^2}{2d}\left((1-r)\varepsilon_0 + r\varepsilon_N(q)\right),
		\end{equation}
		where we introduce $r\equiv l/L$. 
		We note that the parameters $d$ and $r$ are, respectively, the manipulation parameters in this case. 
		We here consider a weak nonlinear permittivity as $\varepsilon_N(q)\simeq \varepsilon_{N0}+\varepsilon_{N1}q^2/2$ taking into account for the symmetry against $q$. 
		Then, the potential can be written as the quartic form 
		\begin{equation}
			U(q,a,b) = \frac{a}{2}q^2 + \frac{b}{4}q^4,
		\end{equation}
		where we rewrite the manipulation parameters as $a\equiv A((1-r)\varepsilon_0 + r\varepsilon_{N0})/d$ and $b\equiv A\varepsilon_{N1}/d$. 
		We note that the work defined by Eq.~{(\ref{eq:work})} corresponds to the mechanical work to change the distance between plates or to insert the medium. 

	\section{Derivations of the main results}\label{app:derivations}
		In this appendix, we show the detailed calculation for the derivation of the main results~{(\ref{eq:wqp_formula})}, {(\ref{eq:wqp_formula_pwr})}, and {(\ref{eq:power})}. 
		The equation of motion is given by 
		\begin{equation}
			\frac{d  q}{dt} = -a  q -b  q^3 +   \xi,
		\end{equation}
		where we substitute the explicit form of the weak quartic potential~{(\ref{eq:potential})} into Eq.~{(\ref{eq:Langevin_NG})}. 
		We assume that $b$ is proportional to a small parameter $\epsilon$, 
		and we expand the solution as $  q(t) =   q_0(t) +   q_1(t)+\dots$, where $  q_0(t) = O(1)$ and $  q_1(t) = O(\epsilon)$. 
		For simplicity, we set the initial condition as $  q(0)=0$. 
		$  q_0$ and $  q_1$ satisfy the following equations: 
		\begin{align}
			\frac{d  q_0}{dt} &= -a  q_0 +   \xi\\
			\frac{d  q_1}{dt} &= -a  q_1 - b  q_0^3,
		\end{align}
		whose solutions are given by 
		\begin{align}
			  q_0(t) &=  \int_0^t dt'\exp{\left[-\int_{t'}^t dsa(s)\right]}  \xi(t')\\
			  q_1(t) &= -\int_0^t dt'\exp{\left[-\int_{t'}^t dsa(s)\right]}b(t')  q_0^3(t').
		\end{align}
		
		\subsection{Work along quasistatic processes}
			We derive the work formula~{(\ref{eq:wqp_formula})} for quasistatic processes. 
			The work for quasistatic processes is given by 
			\begin{equation}
				dW_{\mr{qs}} = -\frac{\langle  q^2\rangle_{\mr{ss}}^{a,b}}{2}da - \frac{\langle  q^4\rangle_{\mr{ss}}^{a,b}}{4}db,
			\end{equation}
			where $\langle \cdot \rangle_{\mr{qs}}^{a,b}$ denotes the average in the steady state under fixed parameters $a$ and $b$. 
			The steady average of $  q^2$ is given by 
			\begin{widetext}
				\begin{align}
					\langle  q^2\rangle_{\mr{ss}}^{a,b} 	&= 	\lim_{t\rightarrow \infty}
																\Bigg[\int_0^t\prod_{i=1}^2ds_i e^{-a(t-s_i)}\langle  \xi_1  \xi_2\rangle 
																-2b\int_0^t\prod_{i=1}^2 ds_ie^{-a(t-s_1)}\int_0^{s_2}\prod_{j=3}^{5}e^{-1(s_2-s_j)}
																\langle \xi_1 \xi_3 \xi_4 \xi_5\rangle \Bigg] + O(\epsilon^2)\notag\\
															&= 	\frac{T}{a} - \frac{3bT^2}{a^3} -\frac{bK_4}{4a^2} + O(\epsilon^2),
				\end{align}
				where we have introduced the notation $  \xi_i \equiv   \xi(s_i)$ and used a relation for the fourth moment~{\cite{Gardiner,Kanazawa2}}
				\begin{equation}
					\langle \xi_1 \xi_3 \xi_4 \xi_5\rangle = 4T^2[\delta(s_1-s_3)\delta(s_4-s_5)+\delta(s_1-s_4)\delta(s_3-s_5)+\delta(s_1-s_5)\delta(s_3-s_4)]
					+ K_4\delta_4 (s_1,s_3,s_4,s_5).
				\end{equation}
			\end{widetext}
			The steady average of $  q^4$ is given by 
			\begin{align}
				\langle  q^4\rangle_{\mr{ss}}^{a,b} 	&= 	\lim_{t\rightarrow \infty}
															\Bigg[\int_0^t\prod_{i=1}^4ds_i e^{-a(t-s_i)}\langle  \xi_1  \xi_2  \xi_3  \xi_4\rangle 
															\Bigg] + O(\epsilon)\notag\\
														&= 	\frac{3T^2}{a^2} + \frac{K_4}{4a} + O(\epsilon).
			\end{align}
			Then, we obtain 
			\begin{align}
				dW_{\mr{qs}} &= \left(\!\!-\frac{T}{2a}\!+\!\frac{3bT^2}{2a^3}\!+\!\frac{bK_4}{8a^2}\!\!\right)da \!-\! \left(\!\frac{3T^2}{4a^2}\!+\!\frac{K_4}{16a}\!\right)db\!+\!O(\epsilon^2)\notag\\
							&= -d\left(\frac{T}{2}\log{a} + \frac{3bT^2}{4a^2} + \frac{bK_4}{16a}\right) + \frac{bK_4}{16a^2}da+ O(\epsilon^2), 
			\end{align}
			which implies Eq.~{(\ref{eq:wqp_formula})}.

		\subsection{Power along slow operational processes}
			We next derive the power formula for slow operational processes~{(\ref{eq:wqp_formula_pwr})} and its optimal protocol and power
			{(\ref{eq:opt_prtcl1}-\ref{eq:power})}.  
			We assume that the speed of the parameters' control is finite but slow: $1/\tau =O(\epsilon)$. 
			Let us introduce scaled parameters $\tl{a}(\tl{s})\equiv a(\tau\tl{s})$ and $\tl{b}(\tl{s})\equiv b(\tau\tl{s})$ with the total operation time $\tau$. 
			In a perturbative calculation with respect to $\epsilon\sim 1/\tau$, 
			$q_0(\tau \tl{s})$ can be expanded as
			\begin{align}
				&  q_0(\tau \tl{s}) 
				= \tau\int_0^{\tl{s}} d\tl{s}'\exp{\left[-\tau\int_{\tl{s}'}^{\tl{s}} d\tl{s}''\tl{a}(\tl{s}'') \right]}  \xi(\tau\tl{s}')\notag\\
				=&  \tau\!\!\int_0^{\tl{s}} \!\!\!\!d\tl{s}'e^{-\tau\tl{a}(\tl{s})(\tl{s}-\tl{s}')}
					\!\!\left[1 \!+\! \tau\frac{(\tl{s}\!-\!\tl{s}')^2}{2}\frac{d\tl{a}(\tl{s})}{d\tl{s}}\right]   \xi(\tau\tl{s}') \!+\! O(\epsilon^2),\label{eq:q_0_fs}
			\end{align}
			where we have used the relation $|\tl{s}-\tl{s}'|\sim 1/\tau$ and 
			\begin{align}
				&\exp{\left[-\tau\int_{\tl{s}'}^{\tl{s}} d\tl{s}''\tl{a}(\tl{s}'') \right]} \notag\\
				=&\exp{\left[\!-\tau\int_{\tl{s}'}^{\tl{s}}\!\! d\tl{s}''\!\!\left\{\!a(\tl{s})\!+\!\frac{d\tl{a}(\tl{s})}{d\tl{s}}\!(\tl{s}''\!-\!\tl{s})
					\!+\!O\left((\tl{s}''\!-\!\tl{s})^2\right)\!\right\}\!\right]\!}\notag\\
				=&\exp{\left[\!-\!\tau(\tl{s}\!-\!\tl{s}')\tl{a}(\tl{s})\!+\!\tau\frac{(\tl{s}\!-\!\tl{s}')^2}{2}\frac{d\tl{a}(\tl{s})}{d\tl{s}}\!
					+\!\tau O\!\left((\tl{s}\!-\!\tl{s}')^3\right)\!\right]}\notag\\
				=&e^{-\tau\tl{a}(\tl{s})(\tl{s}-\tl{s}')}\left[1+\tau\frac{(\tl{s}-\tl{s}')^2}{2}\frac{d\tl{a}(\tl{s})}{d\tl{s}}\right]+O(1/\tau^2). 
			\end{align}
			From a similar calculation, $  q_1(\tau\tl{s})$ is also expanded as 
			\begin{align}
				q_1(\tau\tl{s}) 
				=& -\int_0^{\tau\tl{s}} dt'\exp{\left[-\int_{t'}^t dsa(s)\right]}b(t')  q_0^3(t')\notag\\
				=&-\tau^4\int_0^{\tl{s}} d\tl{s}_1e^{-\tau\tl{a}(\tl{s})(\tl{s}-\tl{s}_1)}b(\tl{s}_1)\notag\\
				&\times	\int_0^{\tl{s}_1}\prod_{i=2}^4 d\tl{s}_ie^{-\tau\tl{a}(\tl{s_1})(\tl{s}_1-\tl{s}_i)}\xi(\tau \tl{s}_i)+O(\epsilon^2).
				\label{eq:q_1_fs}
			\end{align}
			From Eqs.~{(\ref{eq:q_0_fs})} and~{(\ref{eq:q_1_fs})}, we obtain 
			\begin{align}
				\langle   q^2(\tau\tl{s})\rangle &= \frac{T}{\tl{a}} - \frac{3bT^2}{\tl{a}^3} -\frac{bK_4}{4\tl{a}^2} 
													+ \frac{T}{2\tau \tl{a}^3}\frac{d\tl{a}}{d\tl{s}} + O(\epsilon^2),\label{eq:finite1_calc}\\
				\langle   q^4(\tau\tl{s})\rangle &= \frac{3T^2}{\tl{a}^2} + \frac{K_4}{4\tl{a}} + O(\epsilon). \label{eq:finite2_calc}
			\end{align}
			Therefore, we obtain Eqs.~{(\ref{eq:wqp_formula_fs})} and~{(\ref{eq:wqp_formula_fs2})}.
		
			We next consider the rectangular protocol shown in Fig.~{\ref{fig:protocol}} 
			assuming that the arrival time at P$_i$ is given by $\tl{\tau}_i=i/4$ for $i=1,2,3$.
			The optimal scaled protocol $\tl{C}$ is given by the variational principle as follows. 
			We first introduce the Lagrangian $\mc{L}(\tl{a},d\tl{a}/d\tl{s}) \equiv (d\tl{a}/d\tl{s})^2/\tl{a}^3$. 
			Then, the variational principle $\delta S[\tl{C}] = 0$ gives 
			\begin{equation}
				\frac{\partial \mc{L}}{\partial (d\tl{a}/d\tl{s})}\frac{d\tl{a}}{d\tl{s}} - \mc{L} = c^2,
			\end{equation}
			which is equivalent to 
			\begin{equation}
				\frac{1}{\tl{a}^3(\tl{s})}\left(\frac{d\tl{a}(\tl{s})}{d\tl{s}}\right)^2 = c^2,
			\end{equation}
			where $c^2$ is a time-independent constant. 
			Then, we obtain 
			\begin{align}
				\frac{1}{\tl{a}^{3/2}(\tl{s})}\frac{d\tl{a}(\tl{s})}{d\tl{s}} = c,
			\end{align}
			for $0\leq \tl{s}\leq 1/4$, 
			which is equivalent to  
			\begin{equation}
				\tl{a}(\tl{s}) = \left|\frac{4\tl{s}}{\sqrt{a_1}} + \frac{1-4\tl{s}}{\sqrt{a_0}}\right|^{-2}, 
			\end{equation}
			under the condition of $\tl{a}(0) = a_0$ and $\tl{a}(1/4)=a_1$. 
			From a parallel calculation, we obtain 
			\begin{equation}
				\tl{a}(\tl{s}) = \left|\frac{3-4\tl{s}}{\sqrt{a_1}} + \frac{4\tl{s}-2}{\sqrt{a_0}}\right|^{-2},
			\end{equation}
			for $1/2\leq \tl{s}\leq 3/4$, $\tl{a}(1/2) = a_1$ and $\tl{a}(3/4)=a_0$. 
			Equation~{(\ref{eq:wqp_formula_pwr})} predicts that the processes $\rm P_1\rightarrow P_2$ $(1/4\leq \tl{s} \leq 1/2)$ and $\rm P_3\rightarrow P_0$ $(3/4\leq \tl{s} \leq 1)$ 
			are irrelevant for $S[\tl{C}]$ and, therefore, their explicit forms are arbitrary if the assumptions of $\tl{b}(1/4)=b_0$, $\tl{b}(1/2)=0$, $\tl{b}(3/4)=0$, $\tl{b}(1)=b_0$, and $d\tl{b}/d\tl{s}=O(\epsilon)$ are satisfied. 
			Thus, the following process is an optimal protocol for $\tl{b}(\tl{s})$: 
			\begin{equation}
				\tl{b}^*(\tl{s}) = 
					\begin{cases}
						b   & (0\leq \tl{s} \leq \frac{1}{4})\cr 
						2b(1-2\tl{s})   & (\frac{1}{4} \leq \tl{s} \leq \frac{1}{2})\cr 
						0   & (\frac{1}{2} \leq \tl{s} \leq \frac{3}{4})\cr
						b(4\tl{s}-3)   & (\frac{3}{4} \leq \tl{s} \leq 1)
					\end{cases}.
			\end{equation}			
			For this optimal protocol $C_{\rm opt}$, we obtain 
			\begin{equation}
				S[C_{\rm opt}] = 8T\left|\frac{1}{\sqrt{a_0}}-\frac{1}{\sqrt{a_1}}\right|^2, 
			\end{equation}
			which implies Eqs.~{(\ref{eq:ineq_opt})} and~{(\ref{eq:power})}.

	\section{Weakly non-Gaussian noises with an arbitrary potential}\label{app:arbtrary_potential}
		In this appendix, we consider weakly non-Gaussian cases with an arbitrary potential $U(q,\vec a)$ and obtain a work formula along quasistatic processes. 
		We assume that higher-order coefficient $K_{2n}$ in the Kramers-Moyal expansion satisfies $K_{2n}=O(\epsilon)$ for $n\geq 2$ with a small parameter $\epsilon$. 
		The Kramers-Moyal expansion of this system~{\cite{Gardiner}} is given by 
		\begin{equation}
			\frac{\partial P(q,t)}{\partial t} = \frac{\partial }{\partial q}\left[\frac{\partial U(q,\vec a)}{\partial q} 
			+ \sum_{i=1}^\infty \frac{K_{2i}}{(2i)!}\frac{\partial^{2i} }{\partial q^{2i}}\right]P(q,t). 
		\end{equation}
		Let us consider the stationary distribution by the perturbation with respect to $\epsilon$. 
		We expand the stationary distribution as $P_{\rm SS}(q) = P_0(q) + P_1(q)+\dots$, where $P_0(q)=O(1)$ and $P_1(q)=O(\epsilon)$. 
		Then, $P_0(q)$ and $P_1(q)$ satisfy the following equations: 
		\begin{align}
			\frac{\partial U}{\partial q}P_0(q) + T\frac{dP_0(q)}{dq} &= 0\\
			\frac{\partial U}{\partial q}P_1(q) + T\frac{dP_1(q)}{dq} &= -\sum_{i=2}\frac{K_{2i}}{(2i)!}\frac{\partial^{2i-1} }{\partial q^{2i-1}}P_0(q),
		\end{align}
		whose solutions are, respectively, given by
		\begin{align}
			P_0(q) &= \frac{e^{-U(q,\vec a)/T}}{\int_{-\infty}^{\infty} dq'e^{-U(q',\vec a)/T}}\\
			P_1(q) &= P_0(q)\left[C + \sum_{i=2}^\infty \frac{K_{2i}}{(2i)!}\mathcal{U}_{2i}(q)\right].
		\end{align}
		Here, $C$ is a normalization constant satisfying $\int_{-\infty}^\infty dqP_1(q)=0$, and we have introduced 
		\begin{equation}
			\mathcal{U}_{2i} (q) \equiv -\int_{0}^{q} \frac{dq'}{T}e^{\frac{U(q',\vec a)}{T}}\frac{\partial^{2i-1}}{\partial q'^{2i-1}}e^{-\frac{U(q',\vec a)}{T}}.
		\end{equation}
		Then, in the first order perturbation, we obtain an integrated work formula for a quasistatic protocol $C_{\rm qs}$:
		\begin{equation}
			\oint_{C_{\rm qs}} dW = \sum_{i=2}^\infty\frac{K_{2i}}{(2i)!}\oint_{C_{qs}} d\vec a \cdot \vec F^{(2i)}(\vec a)\neq 0,
		\end{equation}
		where 
		\begin{equation}
			\vec F^{(2i)} (\vec a) = \left<\! \frac{\partial U(  q,\vec a)}{\partial \vec a}\mc{U}_{2i}(  q,\vec a) \!\right>_{\mr{eq}}\!\!\!\!\!.
		\end{equation}
		This formula implies that we can extract the work from the non-Gaussian properties of the noise. 
			
	\section{The method of integrating factors}\label{app:integrating_factor}
		We have shown that the integrated quasi-static work is not a scalar potential in general. 
		Here we demonstrate that we can construct a scalar potential by the method of integrating factor, and obtain an inequality similar to the second law
		only in the case with the weakly quartic potential. 
		Integrating factors allow an inexact differential to become an exact differential. 
		For example, in the case of equilibrium thermodynamics, temperature is introduced as the integrating factor for heat~{\cite{Caratheodory,Callen}}. 
		It is known that integrating factors always exist for the case of two parameters.  
		In the present case, we find an integral factor $1/T^*\equiv 1+bK_4/8aT$ in the perturbation with respect to $\epsilon$, and we obtain a thermodynamic scalar potential as 
		\begin{equation}
			G(a,b) \equiv \int \frac{dW_{\mr{qs}}}{T^*} = -\frac{T}{2}\log{a} - \frac{3T^2b}{4a^2} - \frac{bK_4}{16a} +O(\epsilon^2).
		\end{equation}
		Furthermore, for the slow operational processes with $d\tl{a}/ds = O(1)$ and $d\tl{b}/ds = O(\epsilon)$, we can show the following equality
		\begin{equation}
			\int \frac{\langle dW\rangle}{T^*} - G(a,b) = 
			- \frac{1}{\tau}\int_0^1 
			\frac{d\tl{s}T}{4\tl{a}^3}\left[\frac{d\tl{a}}{d\tl{s}}\right]^2 + O(\epsilon^2), \label{eq:wqp_formula_fs_sp}
		\end{equation}
		which implies an inequality similar to the second law as
		\begin{equation}
			\int \frac{\langle dW\rangle}{T^*} \leq G(a,b) + O(\epsilon^2). \label{ineq:tsp}
		\end{equation}
		We note that we obtain such an inequality similar to the second law only for the weakly quartic potential and the slow processes. 
		However, it is unclear whether we can show second-law-like inequalities using the method of integrating factor for general cases. 
		
		We here briefly present the derivation of Eq.~{(\ref{eq:wqp_formula_fs_sp})}. 
		On the conditions of $d\tl{a}/ds = O(1)$ and $d\tl{b}/ds = O(\epsilon)$, we obtain 
		\begin{align}
			&\frac{d\la W\ra}{d\tl{s}} 	= \left(-\frac{T}{2\tl{a}}+\frac{3\tl{b}T^2}{2\tl{a}^3}+\frac{\tl{b}K_4}{8\tl{a}^2}\right)\frac{d\tl{a}}{d\tl{s}} \notag\\
										&- \left(\frac{3T^2}{4\tl{a}^2}+\frac{K_4}{16\tl{a}}\right)\frac{d\tl{b}}{d\tl{s}} - \frac{T}{4\tau \tl{a}^3}\left(\frac{d\tl{a}}{d\tl{s}}\right)^2 + O(\epsilon^2),
		\end{align}
		where we used Eqs.~{(\ref{eq:finite1_calc})} and {(\ref{eq:finite2_calc})}.
		Then, we obtain
		\begin{align}
			\frac{1}{T^*}\frac{d\la W\ra}{d\tl{s}}
			=& \left(-\frac{T}{2\tl{a}}+\frac{3\tl{b}T^2}{2\tl{a}^3}+\frac{\tl{b}K_4}{16\tl{a}^2}\right)\frac{d\tl{a}}{d\tl{s}} \notag\\
			&- \left(\frac{3T^2}{4\tl{a}^2}+\frac{K_4}{16\tl{a}}\right)\frac{d\tl{b}}{d\tl{s}} - \frac{T}{4\tau \tl{a}^3}\left(\frac{d\tl{a}}{d\tl{s}}\right)^2 + O(\epsilon^2)\notag\\
			=& \frac{dG(a,b)}{d\tl{s}} - \frac{T}{4\tau \tl{a}^3}\left(\frac{d\tl{a}}{d\tl{s}}\right)^2 +O(\epsilon^2),
		\end{align}
		which implies Eq.~{(\ref{eq:wqp_formula_fs_sp})}.

\end{document}